\documentclass[11pt,a4paper]{llncs}
\usepackage{amsmath}
\usepackage{amssymb}
\usepackage{graphicx}
\usepackage{marvosym}
\usepackage{url}
\usepackage{fancyhdr}
\usepackage{enumitem}
\usepackage{placeins}
\usepackage{comment}
\usepackage{bm}
\usepackage{booktabs}

\usepackage[style=ieee]{biblatex}  
\usepackage{geometry}
\newcommand{\keywords}[1]{\par\addvspace\baselineskip
\noindent\keywordname\enspace\ignorespaces#1}

\graphicspath{{Figures/}}

\fancypagestyle{firstpage}{\fancyhf{}
}

\begin{document}


\title{\LARGE{Privacy-Preserving Multimodal News Recommendation through Federated Learning}}


%
%

\author{\large{Mehdi Khalaj \and Shahrzad Golestani Najafabadi \and Julita Vassileva}}
\institute{\large{Department of Computer Science, University of Saskatchewan, Canada}\\
\texttt{\{mehdi.khalaj, shg550, julita.vassileva\}@usask.ca}}

\maketitle

\thispagestyle{firstpage}

\begin{abstract}
Personalized news recommendation systems (PNR) have emerged as a solution to information overload by predicting and suggesting news items tailored to individual user interests. However, traditional PNR systems face several challenges, including an overreliance on textual content, common neglect of short-term user interests, and significant privacy concerns due to centralized data storage. This paper addresses these issues by introducing a novel multimodal federated learning-based approach for news recommendation. First, it integrates both textual and visual features of news items using a multimodal model, enabling a more comprehensive representation of content. Second, it employs a time-aware model that balances users' long-term and short-term interests through multi-head self-attention networks, improving recommendation accuracy. Finally, to enhance privacy, a federated learning framework is implemented, enabling collaborative model training without sharing user data. The framework divides the recommendation model into a large server-maintained news model and a lightweight user model shared between the server and clients. The client requests news representations (vectors) and a user model from the central server, then computes gradients with user local data, and finally sends their locally computed gradients to the server for aggregation. The central server aggregates gradients to update the global user model and news model. The updated news model is further used to infer news representation by the server. To further safeguard user privacy, a secure aggregation algorithm based on Shamir's secret sharing is employed. Experiments on a real-world news dataset demonstrate strong performance compared to existing systems, representing a significant advancement in privacy-preserving personalized news recommendation.
\end{abstract}

\section{Introduction}
According to recent studies by Reuters Institute in 2024, 77\% of users access news through web-based platforms, such as social media networks (e.g., Facebook, Twitter) and news aggregation sites (e.g., Microsoft News)\footnote{Reuters Institute 2024 Digital News Report: https://www.digitalnewsreport.org/}. Rapid growth of online news content has led to an information overload problem, necessitating the development of personalized news recommendation systems (PNR) that provide tailored news updates \cite{adomavicius2005toward}  by recommending the most relevant and interesting news to users based on their browsing history. These systems strive to improve user experience by learning the characteristics, attention patterns, and preferences of users to offer personalized recommendations \cite{karimi2018news}. Recent research has focused on using machine learning techniques to extract features and create vector representations of both news articles and user preferences \cite{an2019neural, wu2019neural_a, qi2021hierec, wang2018dkn, zhu2019dan}. 

Despite significant advancements in the field of news recommendation, several challenges remain:

\begin{enumerate}

\item Existing PNR systems predominantly rely on textual content, ignoring visual information, e.g. cover images (Figure 1). However, research has shown that the combination of both textual and visual information influences the likelihood of clicking on news articles \cite{ xun2021we}. 

\item Existing models often focus mainly on long-term user interests, overlooking recent preferences.  This can negatively impact user experience because users' clicks are influenced both by their long-lasting interests \cite{li2014modeling} and by currently trending news.  Therefore, a model that integrates both long-term and short-term interest modeling is required to provide a more accurate user representation. 

\item Conventional PNR methods raise significant privacy concerns due to the centralized storage of users’ historical click data for model training \cite{shin_privacy_2018}. While some users are unwilling to share their data with a central entity, others may feel apprehensive or concerned even when they choose to do so. Moreover, due to the growing enforcement of data protection regulations such as GDPR\footnote{https://gdpr.eu/} and CCPA\footnote{https://oag.ca.gov/privacy/ccpa}, centralized storage and analysis of user data may become increasingly restricted or even impractical in the future \cite{voigt2017eu, albrecht2016gdpr}. It is essential to mitigate the risk of compromising user privacy by applying a privacy-preserving method for news recommendation model training. While centralized learning can achieve high accuracy, a federated learning model that offers comparable performance while preserving user privacy is generally preferred.

\begin{figure*}[t]
    \centering
    \includegraphics[width=12cm,height=4cm]{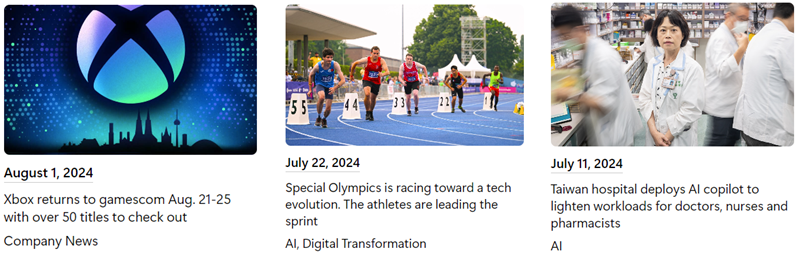}
    \caption{Multimodal News Information (Microsoft News)}
    \label{fig:Figure1}
\end{figure*}

\end{enumerate}

To address key limitations in personalized news recommendation systems, we propose a comprehensive and privacy-preserving framework that tackles three critical challenges in a unified manner: incorporating multimodal content, modeling both short-term and long-term user interests, and ensuring user data privacy. Our approach is motivated by the growing demand for more expressive content representation, time-aware personalization, and privacy-conscious system design. By integrating these components into a cohesive architecture, our model is designed to enhance recommendation quality while aligning with modern privacy standards.

Unlike most prior methods that rely solely on textual information such as news titles, our model enriches input representation by integrating both textual and visual features, including cover images. It further enhances personalization by capturing the temporal dynamics between short-term and long-term user preferences, outperforming approaches that consider only long-term interests. To protect user data, we implement an efficient privacy-preserving framework based on federated learning principles. The effectiveness of this integrated approach—designed to address multiple key challenges—is evaluated against state-of-the-art methods, with a focus on enhancing recommendation accuracy while preserving user privacy.  The main contributions of this work include: 
\begin{itemize}
\item A novel multimodal news model (encoder) that combines textual and visual elements for improved news representation (vector) learning.
\item By combining short-term and long-term user interests and effectively capturing their temporal correlations, our model achieves improved performance over methods that rely solely on long-term interests.
\item An efficient privacy-preserving framework based on federated learning approach to ensure users' privacy in news recommendations.
\item A secure aggregation algorithm based on Shamir's Secret Sharing is introduced to strengthen privacy protection in federated learning against potential privacy attacks.
\end{itemize}

\section{Background and Related Work}
This section provides an overview of the current work in news recommender systems, multimodal news recommendation, the existing federated learning approaches applied to news recommender systems, and finally, secure multi-party computation.

\subsection{News Recommendations Systems}
Personalized news recommendation has evolved significantly with the advent of deep learning techniques, addressing challenges such as cold-start, sparsity, and lack of personalization faced by traditional filtering methods \cite{khusro2016recommender}. Recent years have seen the development of numerous deep learning-based news recommendation models employing various architectures. These models focus on two key aspects: news modeling and user modeling. 

News modeling is a critical step in personalized news recommendation methods to capture the features of news articles and understand their content. Models like Neural News Recommendation with Personalized Attention (NPA) \cite{wu2019npa} by Wu et al. and Neural News Recommendation with Attentive Multi-View Learning (NAML) by Wu et al. \cite{wu2019neural_a} use Convolutional Neural Network (CNN) \cite{lecun_gradient-based_1998} and attention \cite{ashish2017attention} mechanisms to extract semantic features from headlines and news bodies. Transformer-based architectures \cite{ashish2017attention} and pre-trained language models like Bidirectional Encoder Representations from Transformers (BERT) \cite{kim2022bert} have further enhanced news representation capabilities \cite{wu2019npa, wu2019neural_b}. 

User modeling refers to inferring user interests and preferences from their historical click behavior, which is a key step in a personalized news recommendation system. User modeling has similarly evolved, with methods that focus on aggregating representations of previously clicked news. To capture temporal dynamics in user interests, some models employ recurrent neural networks (RNN) \cite{zhu2019dan} or hybrid approaches that combine long-term and short-term interests \cite{an2019neural}. Advanced techniques like multi-head self-attention and graph neural networks \cite{hu2020graph} have been applied to capture complex user-news interactions. These advances in news and user modeling have significantly improved the accuracy and effectiveness of personalized news recommendations, enabling systems to better capture the nuances of user interests and news content.

\subsection{Multimodal News Recommendation}
Online content-sharing platforms are now rich in various modalities (photos, videos, audio podcasts, etc.). Consequently, the emerging field of multi-modal recommendation has gained significant attention \cite{yu2019vision}, finding applications in diverse areas such as fashion recommendations \cite{chelliah2019principle}. Recent studies have explored multimodal approaches in news recommendation systems, combining textual and visual information to better simulate user behavior \cite{xun2021we, han2022vlsnr}. These methods aim to address the limitations of text-only models by incorporating cover images, which significantly impact user engagement. 

Xun et al. \cite{xun2021we} employed an additive attention mechanism for multimodal fusion, which is the process of combining information from multiple different modalities (e.g., text and images). Han et al. \cite{han2022vlsnr} introduced VLSNR (Vision-Linguistics Coordination Time Sequence-aware News Recommendation), a framework that considers both long-term and short-term user preferences by fusing text and image features. It utilizes the Contrastive Language-Image Pre-Training (CLIP) encoder \cite{radford2021learning} to capture comprehensive news features and user behavior patterns across modalities. However, these frameworks do not prioritize preserving user privacy.

\subsection{Federated Learning}
Existing news recommendation methods mainly rely on centralized storage of user browsing behavior for model training. Centralized storage of user data in news recommendation systems poses significant privacy risks, particularly given the sensitive nature of the information involved. Such data may include users’ reading habits, political affiliations, religious beliefs, lifestyle preferences, interactions (such as likes, shares, and comments), geolocation, and device metadata. If a user’s browsing history is compromised, it can lead to serious violations of user rights \cite{ravi_secrecsy_2019}. The high sensitivity of user behavior to privacy underscores the growing concern that central servers can become targets for data breaches, as highlighted in \cite{li2020federated}. Federated learning (FL) \cite{mcmahan2017communication} can address privacy concerns in news recommendation systems by allowing collaborative model training without centralized data storage. 

Recent studies have explored federated learning in news recommendation systems to enhance privacy protection. Qi et al. \cite{qi2020privacy} proposed FedRec, a privacy-preserving news recommendation framework that trains models locally on user devices with added noise using local differential privacy (LDP) \cite{ren2018textsf}. Each device maintains a full model copy, computes gradients from user behavior, and sends them to a central server for aggregation and global updates. Yi et al. \cite{yi2021efficient} proposed Efficient-FedRec, an optimized framework that partitions the model into two parts: a large news model residing solely on the server, and a smaller user model and news representation module shared between the server and client, significantly reducing the resource burden on user devices.

\subsection{Secure Multi-Party Computation}
Federated learning, while offering significant potential benefits, faces several security and privacy challenges. Recent research has highlighted concerns that can compromise privacy guarantees, leading to increased focus on developing privacy protection methods. Various privacy-enhancing techniques, such as secure multiparty computation (SMC) \cite{knott2021crypten}, Homomorphic Encryption \cite{phong_privacy-preserving_2018}, and local differential privacy \cite{ren_lopub_2018}, have been proposed to prevent private information leakage and mitigate cyber-attack risks in Federated Learning-based frameworks. SMC allows parties to participate in confidential computing without revealing their private data, effectively addressing the conflict between data privacy and sharing \cite{nosouhi_ucoin_2023}.

\section{Methodology}

This section introduces Fed-MM-PNR, a privacy-preserving news recommendation framework that combines textual and visual features to enrich input representation and captures both short-term and long-term user interests for improved personalization. To protect user data, the system leverages federated learning and incorporates a secure aggregation protocol based on secure multi-party computation (SMC) to defend against privacy attacks.

\subsection{News Recommendation Architecture}
In this subsection, we present the news recommendation architecture proposed in this study. The framework consists of three core modules: the news encoder, the user encoder, and the click score predictor.

\subsubsection{Multimodal News Model}
\

The design of an effective news model (encoder) for recommendation systems must consider multiple factors that influence user engagement, including both textual and visual elements of news articles. Recognizing the importance of these diverse information sources, a comprehensive news encoder has been developed to extract information from both the visual and textual components of news items. This approach utilizes multimodal models to capture the semantics of text and images in a shared metric feature space, acknowledging the significant role that images play in attracting user attention and influencing click decisions. The news model architecture (Figure 2) comprises three key components: a text encoder, an image encoder, and a late-fusion model.

\begin{figure}[htp]
    \centering
    \includegraphics[width=10cm,height=8cm]{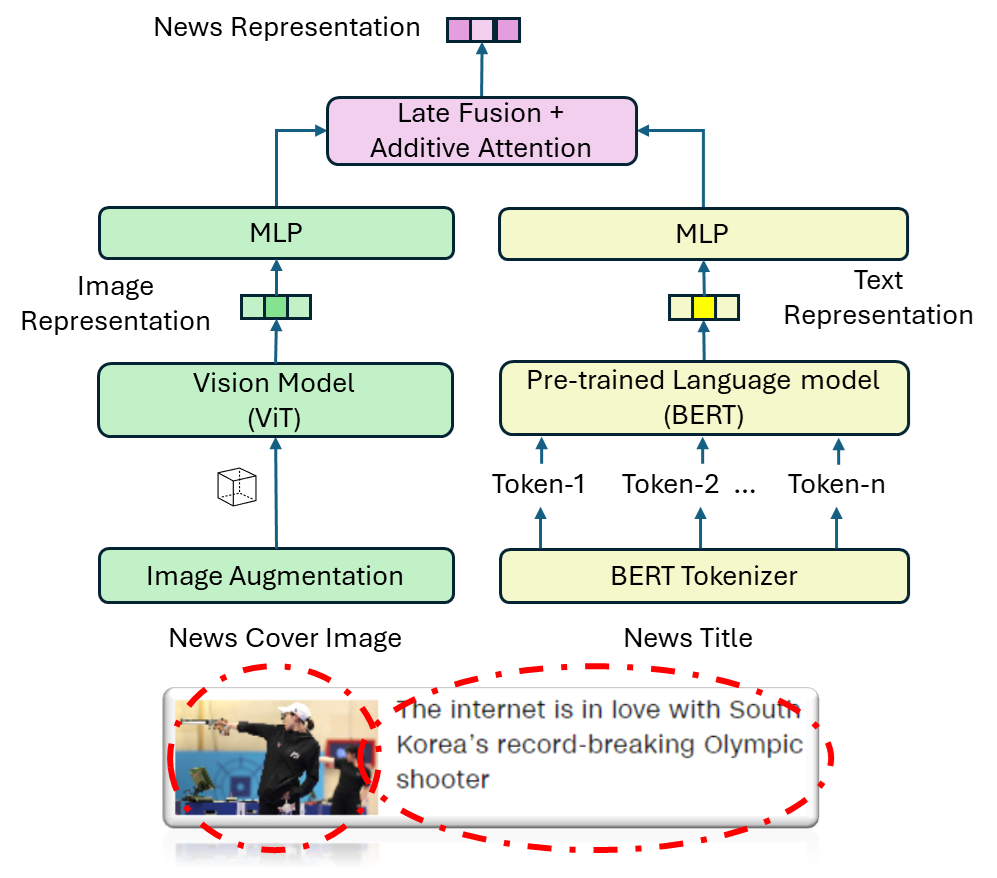}
    \caption{News Encoder Architecture}
    \label{fig:Figure3}
\end{figure}

{\bfseries The text encoder} focuses on processing news titles and consists of three layers. The first layer uses the BERT tokenizer \cite{kim2022bert} to transform titles into word sequences. The second layer incorporates a pre-trained BERT model to generate contextually aware representations of the input text. The final layer is a multilayer perceptron (MLP) that projects the BERT output into a suitable format for the fusion layer. 

{\bfseries The image encoder} processes news cover images in three layers. The initial layer applies various image augmentations such as rotation, resizing, normalizing, horizontal flip, and random brightness/contrast to increase data diversity and model robustness. The second layer utilizes a Vision Transformer (vit\_base\_patch16) \cite{dosovitskiy_image_2021} introduced by Dosovitskiy et al., which is a vision model used to extract rich visual representations. The final layer is an MLP that transforms these representations for compatibility with the fusion layer.

{\bfseries The late fusion model} combines the output of the text and image encoders. This approach processes each modality independently before combining its outputs, allowing for the learning of rich, modality-specific information. The late fusion technique enables the creation of a multimodal representation that captures semantic concepts from unimodal features, resulting in a comprehensive understanding of news content that takes advantage of both textual and visual information. In this project, the late fusion is done by concatenating the image and text features and applying an additive attention mechanism to produce the final news representation.

\subsubsection{User Model}
\

The user model (encoder) in this system learns user representation (vector) by analyzing the user's browsing history and focusing on clicked news articles. It transforms the sequence of news (including both news title and cover image) clicked by the user into news embeddings through a multimodal news encoder and processes these embeddings to generate a comprehensive user embedding. The model captures both long-term and short-term user interests, providing a nuanced understanding of user preferences. The user representation is ultimately formed by combining these two aspects of interest, as illustrated in Figure 3. Denoting the news representations of the user's historically clicked news 
\begin{math}
    \bm{\left[n_1,n_2,\ldots,n_N \right]}
\end{math}
as input, the user model computes the user representations \textbf{\textit{u}} as output. This approach allows for a more dynamic and accurate representation of user behavior, which could lead to more personalized and effective content recommendations.

\begin{figure}[htp]
    \centering
    \includegraphics[width=9cm,height=9cm]{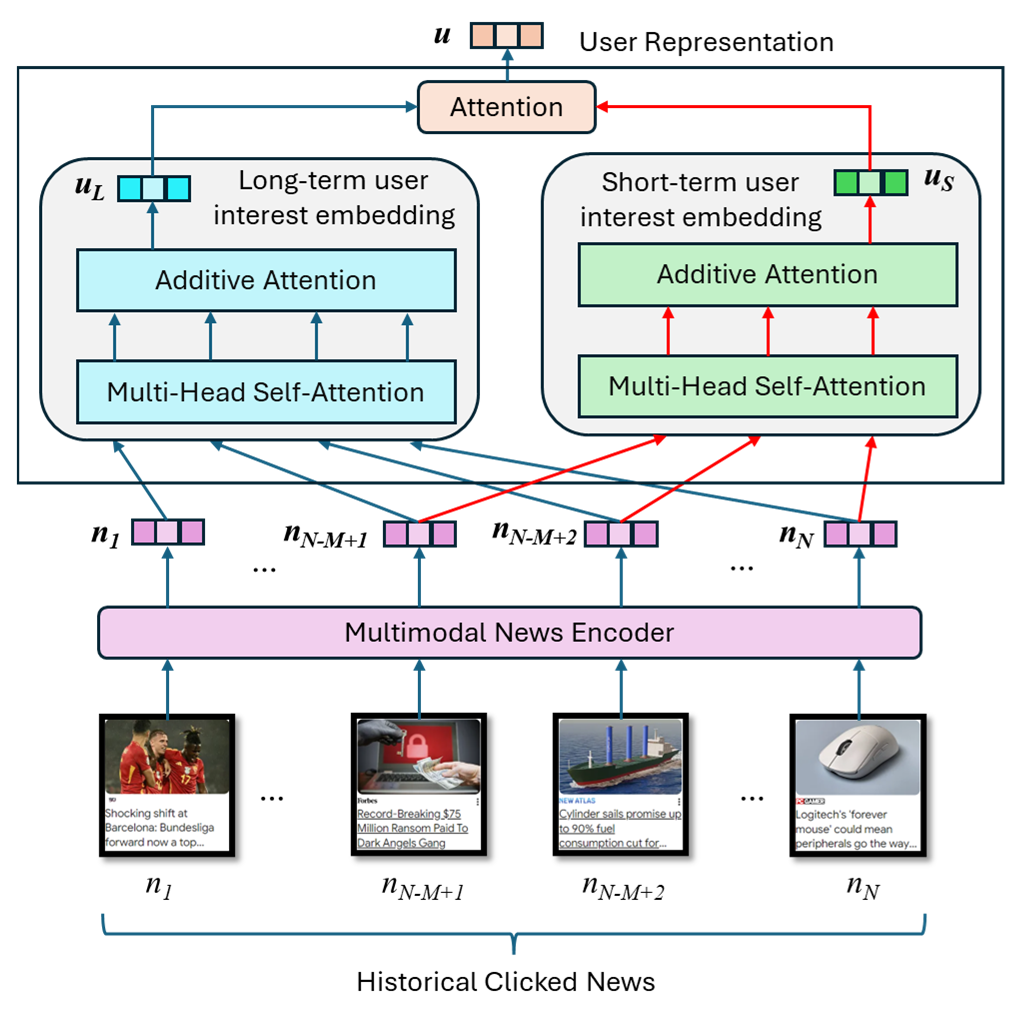}
    \caption{ User Encoder Architecture}
    \label{fig:Figure4}
\end{figure}

{\bfseries Long-term interest modeling} utilizes multi-head self-attention and additive attention networks to capture correlations between different news articles read by the same user, and finally, the long-term representation 
\begin{math}
    \bm{u_L}
\end{math} of the user is obtained. 

{\bfseries Short-term interest modeling} focuses on the most recent batch of news vectors, applying the same network structure (multi-head self-attention and additive) to capture sequential characteristics. If the length of the most recent news sequence to obtain the short-term interest is set to \textit{M}, then the input of the short-term user interest modeling is 
\begin{math}
  \bm{\left[n_{N-M+1},n_{N-M+2},..., n_N\right]}
\end{math}, and finally, the short-term representation 
\begin{math}
  \bm{u_S}
\end{math} of the user is obtained.

This dual approach allows for a more comprehensive understanding of user preferences, addressing both established interests and recent changes in behavior. 
To integrate long-term and short-term user representations, the approach uses an additive attention network that dynamically assigns weights to both long-term and short-term interest representations which reflects their relative importance for each unique user. The output of this attention network serves as the final user representation.

\subsubsection{News Ranking and Model Training}
\

The click score prediction model aims to generate a user's rating for candidate news using the user and candidate news representations derived from the encoder. For each user (\textit{u}), the representations of their historically clicked news are used by the local user encoder to create a user representation vector 
\begin{math}
  \bm{u}
\end{math}. For each candidate news 
\begin{math}
  n_c
\end{math}, its embedding vector is 
\begin{math}
  \bm{n_c}
\end{math}. The final ranking score 
\begin{math}
  s(u,n_c)
\end{math} is defined as the dot product of their embedding vectors: 
\begin{math}
  s(u,n_c)=\bm{u^Tn_c}
\end{math} 
\cite{ni2021effective}.

News recommendation models are trained by analyzing users' clicking behavior on news platforms. When a user clicks on an item, the model increases the ranking score for that specific user-news pair, while a lack of clicks decreases it. The proposed training approach considers both click and non-click behaviors within the same impression. The training uses categorical cross-entropy loss. For every clicked news item by the user \textit{u}, \textit{K} non-clicked news items from the same impression are randomly sampled, forming an input sample of \textit{K+1}
 news items. As referenced in studies like Fed-Rec \cite{qi2020privacy}, the loss for a training sample is calculated based on the total number of user clicks 
 (\begin{math}
    B_u
\end{math}) as follows:
\vspace{-1mm}
\begin{equation}
  L_u^i=-log(\frac{exp(s(u,\ n_i^c))}{exp(s(u,\ n_i^c))+ \sum_{j=1}^{k}exp(s(u,\ n_{i,j}^{nc}))})
\end{equation}

where 
\begin{math}
    n_i^c
\end{math} and 
\begin{math}
    n_{i,j}^{nc} 
\end{math}
are the news clicked and unclicked by users in the same impression respectively, and \textit{s(u,n)} is the ranking score of the news \textit{n} for user \textit{u}.
The final loss function of the news recommendation model is the average loss of all training samples in 
\begin{math}
    B_u
\end{math}, which is computed as follows:
\vspace{-3mm}
\begin{equation}
  L_u=-\frac{1}{\left|B_u\right|}\sum_{j=1}^{\left|B_u\right|}L_u^j
\end{equation}

\subsection{Federated News Recommendation Framework}

\begin{figure*}[t]
  \includegraphics[width=\linewidth,height=6cm]{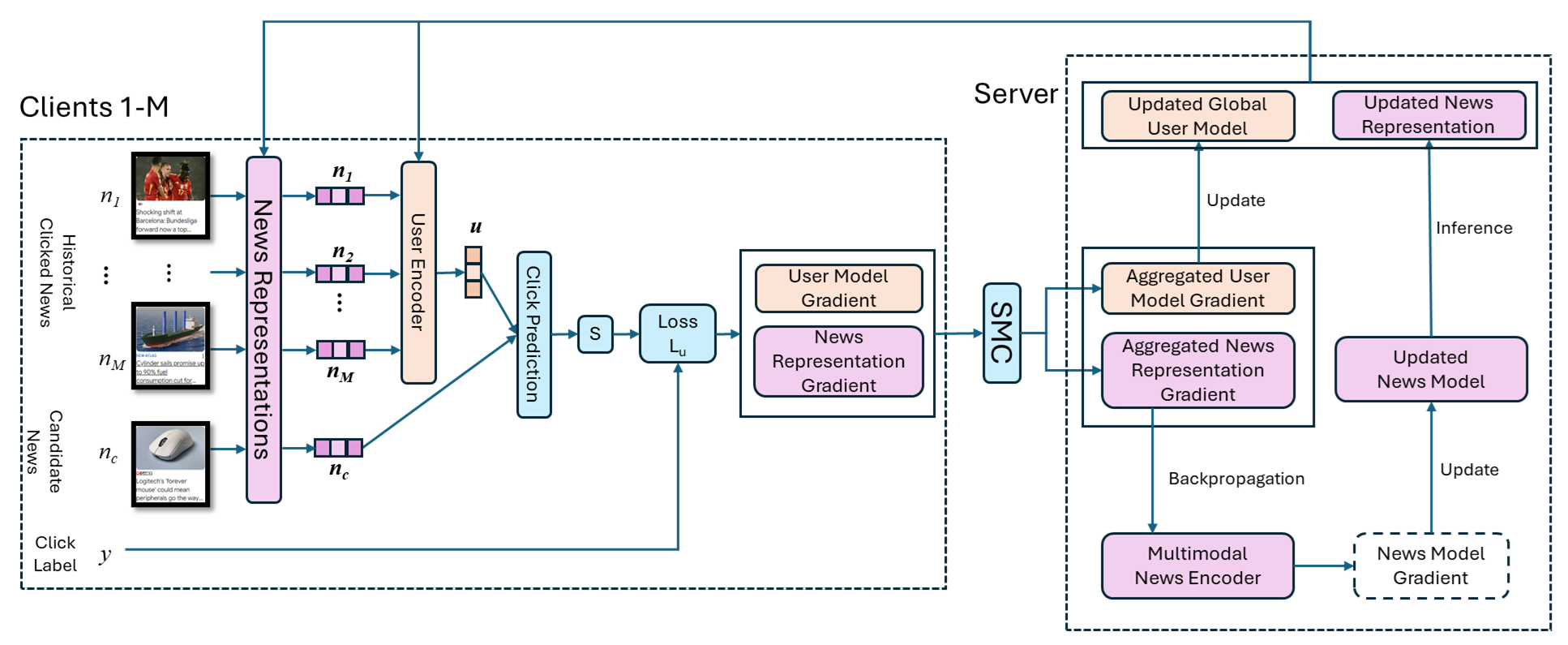}
  \caption{Fed-MM-PNR Framework}
\end{figure*}

In this subsection, we integrate the news recommendation architecture introduced earlier with an efficient federated learning framework as presented in \cite{yi2021efficient} to build Fed-MM-PNR, a personalized and privacy-preserving news recommendation model. The general architecture of the proposed federated news recommendation system is depicted in Figure 4. Unlike traditional centralized systems, Fed-MM-PNR addresses privacy concerns by ensuring that client data never leaves the local device. Instead of uploading raw data, clients train the user model locally using their browsing history and transmit only the local loss gradients to a central server.

To optimize efficiency and reduce communication overhead, Fed-MM-PNR adopts a model decomposition strategy by splitting the full recommendation model into two parts: (1) a large news representation model, which remains on the server, and (2) a lightweight user model, which is shared between the server and the clients. In each federated learning round, a subset of clients is randomly selected to participate in training. Random selection is used because not all clients may be available or capable of contributing at any given time due to factors such as limited connectivity, device constraints, or privacy preferences. During training, each selected client performs local training using their private data and only a small subset of the news representations (vectors) retrieved from the server, rather than downloading the entire news model. Specifically, the client requests only the news representations corresponding to the news articles they have clicked on, minimizing communication overhead and enhancing system efficiency. This design significantly reduces the burden on the client-side, as it limits both the amount of computation and the volume of data that needs to be transmitted.

After retrieving the news representations related to its historically clicked news, each selected client computes the gradients of their local user model, which are then sent to the server. The server aggregates the gradients and uses them to update the global user model and, if necessary, relevant parts of the news model. Updated parameters are subsequently sent back to the clients for the next training iteration. This process repeats iteratively until convergence. By combining personalization, decentralization, and communication efficiency, Fed-MM-PNR enables multiple users to collaboratively train a high-quality recommendation model while preserving their privacy. Furthermore, the decomposition of the model reduces latency and resource consumption on user devices, making Fed-MM-PNR a scalable and practical solution for real-world news recommendation systems where there are hundreds, thousands, or millions of users and privacy and efficiency are critical.

\subsection{Secure Aggregation}
To protect user privacy in model training, we need to develop a secure aggregation protocol using Secure multi-Party Computation (SMC) as the defense mechanism. 
This study employs the Direct Communication Model for enhanced security in federated learning, allowing parties to exchange encrypted data directly without a trusted intermediary. Shamir's Secret Sharing is utilized to divide secrets into multiple shares \cite{shamir1979share}. Participants split their secret data into shares, distribute them among other participants, and collectively compute the aggregated value without exposing individual data. This approach ensures privacy while allowing for effective collaborative learning in a decentralized manner. 

In our framework, the first time we apply Shamir's secret sharing is to calculate the "aggregated news pool" of a set of users and request the news representation of the joint news set from the server to avoid privacy leakage from individual users. This approach hides individual click histories while allowing the server to know only group-level interactions. The technique enables the central server to compute the sum without accessing local client vectors, effectively protecting user privacy while reducing communication costs in federated learning scenarios. In the same way, our framework employs Shamir's Secret Sharing for the second time to securely aggregate gradient updates from multiple users. This approach maintains the privacy of each user's gradient vectors while allowing the server to perform the necessary computations for model updates without being a trusted entity.

For 
\begin{math}
    U_m\ =\left\{u_1, u_2, \ldots, u_M\right\}
\end{math}, the \textit{M} users chosen by the server, each user's local vector is represented as 
\begin{math}
    h_i\ =\ \left[h_{i,1},\ h_{i,2},\ \ldots,h_{i,N}\right]
\end{math}. Our framework utilizes Secure Aggregation with Shamir’s Secret Sharing to generate the "Aggregated news pool (h)" as follows:

Assuming the user 
\begin{math}
    u_i
\end{math} requests a local news set of 
\begin{math}
    N_i
\end{math}, which is converted into a local vector 
\begin{math}
    h_i
\end{math}  whose dimension is equal to the number of all news (\textit{N}). The j-th dimension of 
\begin{math}
    h_i
\end{math} is defined as:
\begin{equation}
    h_{i,j}\ =\ 1,\ if\ nj\ \in\ Ni\ \ and\ 0,\ otherwise
\end{equation}
where 
\begin{math}
    n_j
\end{math} is the 
\begin{math}
    j^{th}
\end{math} news item in the whole news set \cite{yi2021efficient} and 
\begin{math}
    h_i\ =\ [h_{i,1},\ h_{i,2},\ \ldots,h_{i,N}]
\end{math}.

Each user 
(\begin{math}
   u_i
\end{math}) splits each of the elements in 
\begin{math}
   h_i
\end{math} into M (the number of all users) random parts such that 
\begin{math}
   h_{i,j}=\ \sum_{k=1}^{M}h_{i,j}^k
\end{math}. Thus, each user 
(\begin{math}
   u_i
\end{math}) converts the vector 
\begin{math}
   h_i=[h_{i,1},\ h_{i,2},\ \ldots,h_{i,N}]
\end{math} to a matrix with dimension 
\begin{math}
   M\times N 
\end{math} as follows:

\begin{equation}
    H_i\ =\left[\begin{matrix}h_{i,1}^1&\cdots&h_{i,N}^1\\\vdots&\ddots&\vdots\\h_{i,1}^M&\cdots&h_{i,N}^M\\\end{matrix}\right]\ =\ \left[\begin{matrix}h_i^1\\\vdots\\h_i^M\\\end{matrix}\right]_{M\ \times\ N}
\end{equation}
Where 
\begin{math}
    h_i^k\ =\ \left[h_{i,1}^k,h_{i,2}^k,\ldots,h_{i,N}^k\right]
\end{math} and k is the ID of the user 
(\begin{math}
   u_k
\end{math}) that will receive 
\begin{math}
   h_i^k
\end{math} from 
\begin{math}
   u_i
\end{math}.

Then, each user sends these parts to all other users so that 
\begin{math}
   h_i^1 
\end{math} is sent by 
\begin{math}
   u_i
\end{math} to 
\begin{math}
   u_1
\end{math} and so on. It means every user receives parts from every other user, ensuring that each user holds a share from every other user. Each user then stores the parts they receive so that 
\begin{math}
   u_j
\end{math} stores 
\begin{math}
   h_1^j,\ h_2^j,\ ...\ ,\ h_M^j 
\end{math} such that:
\begin{equation} 
    \begin{aligned}    H_j=\left[\begin{matrix}h_1^j\\\vdots\\h_M^j\\\end{matrix}\right]=\left[\begin{matrix}h_{1,1}^j&\cdots&h_{1,N}^j\\\vdots&\ddots&\vdots\\h_{M,1}^j&\cdots&h_{M,N}^j\\\end{matrix}\right]_{M\ \times\ N} \\
    h_i^j= \left[h_{i,1}^j,h_{i,2}^j,\ldots,h_{i,N}^j\right]_{1\ \times\ N}
    \end{aligned}
\end{equation}

Where
\begin{math}
   u_i
\end{math} sends the vector 
\begin{math}
   h_i^j
\end{math} to 
 \begin{math}
   u_j
\end{math} and 
 \begin{math}
   i,j\in\left\{1,..,M\right\}
\end{math}.

Finally, each user sums up the columns of 
\begin{math}
   H_j 
\end{math} to create a vector of size N and send the final vector 
(\begin{math}
   V_j
\end{math}) to the server:
\begin{equation}
    V_j\ =\ \left[\begin{matrix}\sum_{i=1}^{M}h_{i,1}^j&\ldots&\sum_{i=1}^{M}h_{i,N}^j\\\end{matrix}\right]
\end{equation}

The server can now reconstruct the original secret \textit{h} by summing up all received parts, including their own:
\begin{equation}
    h=\ \sum_{j=1}^{M}V_j=\left[\begin{matrix}\sum_{j=1}^{M}{\left(\sum_{i=1}^{M}h_{i,1}^j\right),}&\ldots,&\sum_{j=1}^{M}\left(\sum_{i=1}^{M}h_{i,N}^j\right)\\\end{matrix}\right]_{1\times N}
\end{equation}

The news requested from the server in the end is represented as an aggregated news pool corresponding to a dimension in the vector \textit{h} that is non-zero. This pool, denoted as 
\begin{math}
   N_m=\ \bigcup_{u_i\in U_m} N_i
\end{math} and combines the news accessed by multiple users. Consequently, the server only knows the collective news accessed by a set of clients, not individual browsing histories.

\section{Experiments and Discussion}
\subsection{Experimental Design}
\textit{\textbf{Dataset}.} The only news recommendation datasets that include both the title and the cover image of news articles are MIND \cite{wu2020mind}\footnote{https://msnews.github.io}, IM-MIND \cite{wu2021mm}, and V-MIND \cite{han2022vlsnr}. The original MIND dataset provides a large-scale collection of English news recommendation data, including user interactions over a six-week period. To enhance its multimodal capabilities, IM-MIND and V-MIND extended MIND by associating most news items with corresponding cover images, resulting in a total of 130,379 images to support comprehensive text-image analysis in multimodal research. The MIND dataset is already pre-divided into separate training, validation, and testing sets by default; therefore, no additional data splitting was necessary. The detailed statistical information of the dataset is summarized in Table 1.

\renewcommand{\arraystretch}{0.5}
\begin{table}[h]
\centering
  \caption{Statistics of our Dataset}
  \label{tab:freq}
  \begin{tabular}{ccl}
    \toprule
    Dataset & MIND-Large\\
    \midrule
    \# Users & 876,956\\
    \# News with Images & 130,379\\
    Avg. clicked news & 17.03\\
    Avg. words per line & 4.11\\
  \bottomrule
\end{tabular}
\end{table}

\noindent \textit{\textbf{Parameter Settings}.}
In our experiments, all required parameters were configured as specified in Table 2. We employed negative sampling techniques that combine each positive sample with several negative samples to construct labeled samples for model training. For each clicked sample (regarded as a positive sample), we randomly sampled 20 of the non-clicked ones (regarded as negative samples). Each test was repeated five times.

\renewcommand{\arraystretch}{0.5}
\begin{table}[h]
\centering
  \caption{Parameter Setting}
  \label{tab:freq}
  \begin{tabular}{lcl}
    \toprule
    Parameter & Value\\
    \midrule
    Optimization algorithm  & NAdam\\
    Learning rate  & 6e-5\\
    Batch size & 256\\
    News representations dimension & 400x1\\
    Length of Long-Term News Sequences & 50\\
    Length of Short-Term News Sequences & 20\\
    Maximum news title length & 30 words\\
  \bottomrule
\end{tabular}
\end{table}

\noindent \textit{\textbf{Evaluation metrics}.}
Evaluation metrics included AUC (Area Under the Curve) , MRR (Mean Reciprocal Rank), nDCG@5, and nDCG@10 (Normalized Discounted Cumulative Gain at 5 and 10) to assess the model's performance comprehensively. (a) AUC measures the model's ability to distinguish between positive and negative classes. (b) MRR  evaluates how early the first relevant item appears in the recommendation list by averaging the reciprocal of its rank across all users. (c) nDCG@5 and nDCG@10 assess the ranking quality of the top 5 and top 10 recommended items, giving higher scores to relevant items appearing earlier in the list.

\noindent \textit{\textbf{Baselines}.}
In our experiments, We compared our proposed method with top-performing, state-of-the-art baselines introduced in recent years. These baselines were sourced from previously published studies, and since we did not have access to their original implementations or the ability to fine-tune their hyperparameters, we relied on the performance metrics reported in their respective papers for comparison. 

\begin{itemize}[itemsep=5pt]
\item Unimodal Centralized: (1) DKN \cite{wang2018dkn} embeds news articles using a knowledge-aware CNN and external knowledge graphs. (2) NPA \cite{wu2019npa} employs personalized attention networks with user ID embeddings. (3) NRMS \cite{wu2019neural_b} uses multi-head self-attention networks for user and news representations. (4) LSTUR \cite{an2019neural} captures long-term and short-term user interests through ID embeddings and GRU sequences. (5) DivHGNN \cite{zhang2024heterogeneous} represents the heterogeneous content of news and the heterogeneous user-news relationships as an attributed heterogeneous graph. (6) NWT \cite{pu2024news} jointly models word relevance and topic prediction through a multi-head self-attention network, integrating precise topic information and semantic relationships. (7) Hyper4NR \cite{liu2024dual} leverages a dual-view hypergraph structure to model users' click history, integrating topic-view and semantic-view hyperedges to capture high-order relations among clicked news. 

\item Unimodal Federated: (8) FedRec \cite{qi2020privacy} uses federated learning and local differential privacy for news recommendations. (9) Efficient-FedRec \cite{yi2021efficient} enhances PLM-NR efficiency and privacy. 

\item Multimodal Centralized: (10) NRMS-IM and FIM-IM \cite{xun2021we} use ResNet101 for the extraction of image features. (11) VLSNR \cite{han2022vlsnr} integrates Visiolinguistic news representations with time-sequence modeling. 

\item Multimodal Federated: (12) Our multimodal federated learning framework, Fed-MM-PNR, integrates both textual and visual features of news, capturing long-term and short-term user interests, and preserving user privacy.

\end{itemize}

\renewcommand{\arraystretch}{1.1}
\begin{table*}[t]
  \caption{News Recommendation Performance of Different Methods}
  \label{tab:commands}
  \centering
  \resizebox{\linewidth}{!}{%
  \begin{tabular}{llccccccc}
    \toprule
    Category & Methods & Federated & Multimodal & Long+Short Term & AUC & MRR & nDCG@5 & nDCG@10\\
    \midrule
    Unimodal Centralized & DKN & \textemdash & \textemdash & \textemdash & 0.672 & 0.321 & 0.353 & 0.417 \\
    & NPA & \textemdash & \textemdash & \textemdash & 0.675 & 0.326 & 0.358 & 0.422 \\
    & NRMS & \textemdash & \textemdash & \textemdash & 0.676 & 0.323 & 0.356 & 0.422 \\
    & LSTUR & \textemdash & \textemdash & \checkmark & 0.680 & 0.323 & 0.363 & 0.427 \\
    & DivHGNN & \textemdash & \textemdash & \checkmark & 0.684 & 0.340 & 0.370 & 0.430 \\
    & NWT & \textemdash & \textemdash & \textemdash & 0.685 & 0.336 & 0.372 & 0.434 \\
    & \textbf{Hyper4NR} & \textemdash & \textemdash & \textemdash & \textbf{0.704} & \textbf{0.356} & \textbf{0.388} & \textbf{0.445} \\
    \midrule
    Unimodal Federated & FedRec & \checkmark & \textemdash & \checkmark & 0.672 & 0.328 & 0.354 & 0.422 \\
    & Efficient-FedRec & \checkmark & \textemdash & \textemdash & 0.680 & 0.330 & 0.361 & 0.422 \\
    \midrule
    Multimodal Centralized & NRMS-IM & \textemdash & \checkmark & \textemdash & 0.687 & 0.331 & 0.369 & 0.432 \\
    & FIM-IM & \textemdash & \checkmark & \textemdash & 0.691 & 0.336 & 0.373 & 0.436 \\
    & VLSNR & \textemdash & \checkmark & \checkmark & 0.695 & 0.340 & 0.376 & 0.440 \\
    \midrule
    Multimodal Federated & \textbf{Fed-MM-PNR (Ours)} & \textbf{\checkmark} & \textbf{\checkmark} & \textbf{\checkmark} & \textbf{0.698} & \textbf{0.350} & \textbf{0.384} & \textbf{0.442} \\
    \bottomrule
  \end{tabular}%
  }
\end{table*}

\newpage

\subsection{Performance Evaluation}

\textit{\textbf{Performance of Different Methods}.} 

In this section, we compare the performance of the Fed-MM-PNR framework with several baseline methods, including unimodal centralized, unimodal federated, and multimodal centralized news recommendation approaches. Since baseline methods typically reported the average of five runs, we follow the same approach and report the average over five runs in our experiments, illustrated in Table 3.

Fed-MM-PNR demonstrates strong performance compared to existing unimodal centralized methods, as shown in Table 3. Specifically, it outperforms all unimodal centralized methods except Hyper4NR \cite{liu2024dual} across all four metrics. The superiority of Fed-MM-PNR is attributed to its multimodal architecture, which more effectively captures user preferences by combining textual and visual features. Although Hyper4NR slightly outperforms Fed-MM-PNR, it operates in a centralized setting where user privacy is not preserved. Given that its performance advantage is marginal (0.86\% in AUC, 1.7\% in MRR, 3.2\% in nDCG@5, and 0.7\% in nDCG@10), Fed-MM-PNR presents a compelling alternative, offering comparable accuracy while preserving user privacy through data decentralization.

Fed-MM-PNR also outperforms existing unimodal federated news recommendation methods such as FedRec and Efficient-FedRec, primarily due to its use of multimodality. By integrating news cover images with titles, it more accurately simulates user behaviour through the combination of visual and textual information. Moreover, Fed-MM-PNR enhances user modelling by employing a self-attention network to capture short-term interests and additive attention to integrate both short- and long-term preferences for a comprehensive user representation.

Remarkably, Fed-MM-PNR outperforms baseline multimodal centralized models such as NRMS-IM, FIM-IM, and VLSNR, despite operating under the constraints of federated learning and secure aggregation. While centralized models like VLSNR benefit from access to all the training data from all users in a single location, Fed-MM-PNR processes user data locally, thus preserving user privacy. The Fed-MM-PNR's superior performance can be attributed to a more effective model architecture and the joint optimization of short- and long-term user interests in a multimodal setting. This demonstrates that with careful model design and efficient system coordination, federated learning with privacy guarantees can achieve and even surpass the performance of centralized methods, without compromising user data security. In conclusion, Fed-MM-PNR outperforms most baseline models while effectively preserving user privacy.

\

\noindent \textit{\textbf{Effectiveness of Multimodal Information}.}

To evaluate the impact of multimodal information on news representation effectiveness, we compared Fed-MM-PNR with its single-modality variants—one using only news titles and the other using only cover images—using the AUC and nDCG@10 metrics. As shown in Figure 5, the multimodal approach outperforms both single-modality models across both metrics. Specifically, incorporating both textual and visual features results in more accurate news representations, thereby improving recommendation performance. Notably, the text-only model performs better than the image-only model, but both are surpassed by the multimodal model. 
These results highlight the importance of combining news titles with cover images.

\begin{figure}[htp]
    \centering
    \includegraphics[width=10cm,height=5cm]{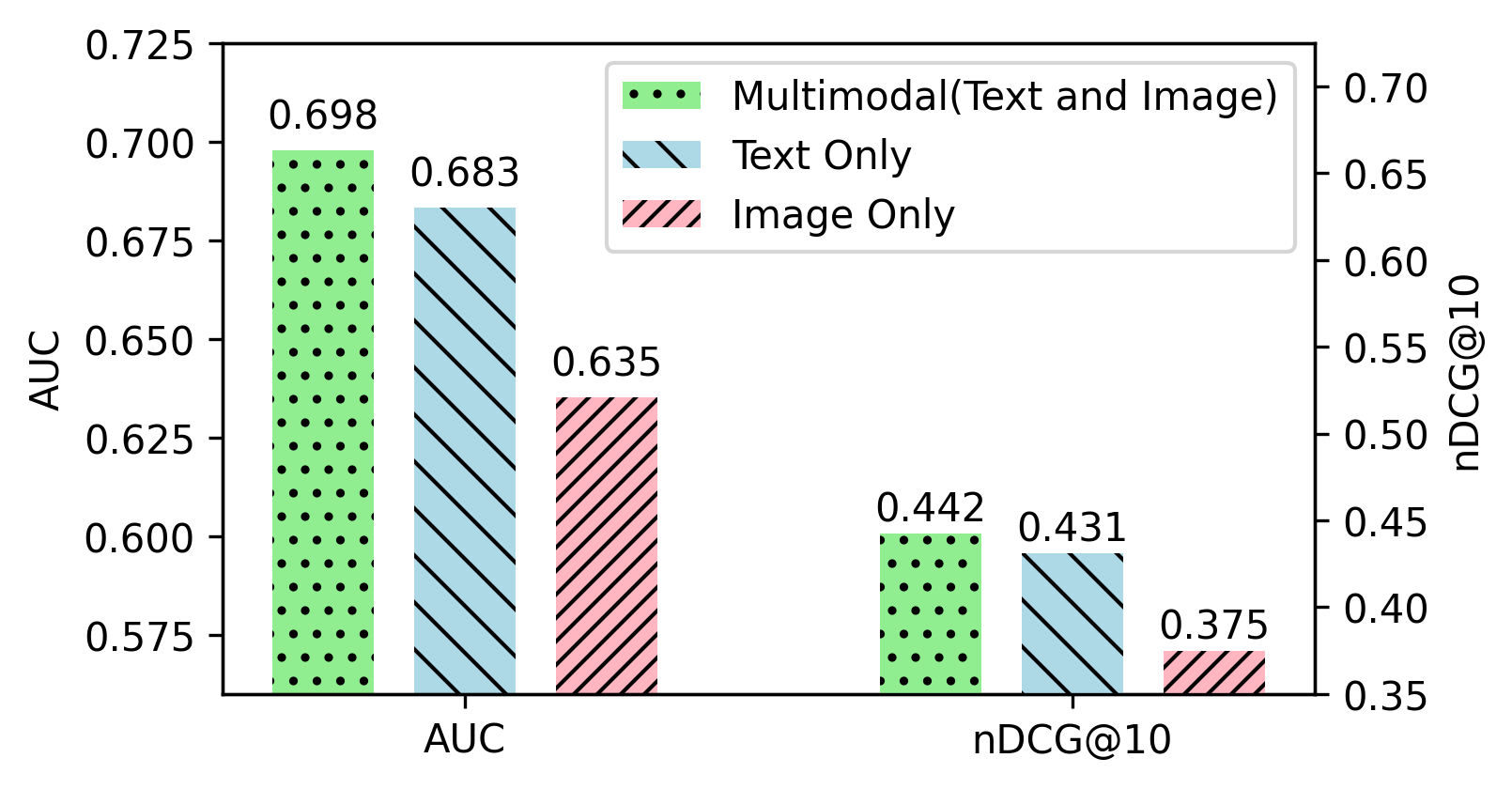}
    \caption{Effectiveness of Multimodal Information}
    \label{fig:Figure6}
\end{figure}

\

\noindent \textit{\textbf{Effectiveness of Combining Short-Term and Long-Term User Interests}.} 

Figure 6 illustrates the impact of incorporating short-term user representations into our news recommendation model, evaluated using AUC and nDCG@10 metrics. The results consistently show that the best performance is achieved when both long-term and short-term user interests are considered. This combined approach outperforms models that rely solely on long-term or short-term interests, with long-term-only models performing better than short-term-only ones. These findings underscore the importance of integrating both temporal aspects of user behavior. The short-term user representation, constructed using a self-attention mechanism, effectively captures recent browsing history, allowing the model to adapt to users’ evolving interests and current events. 

\begin{figure}[htp]
    \centering
    \includegraphics[width=10cm,height=5cm]{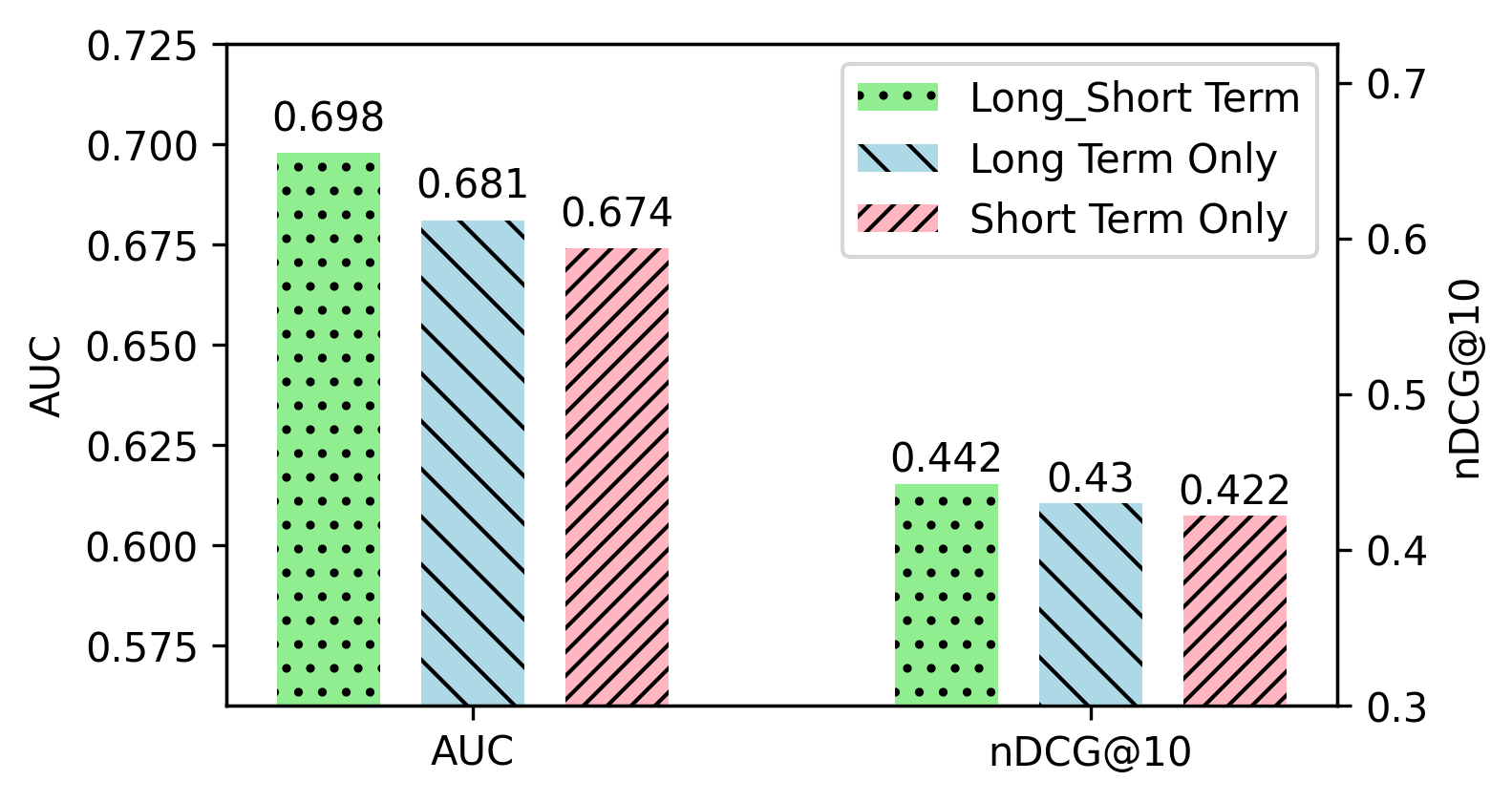}
    \caption{ Effectiveness of Combining Short-Term and Long-Term User Interests}
    \label{fig:Figure7}
\end{figure}

\noindent \textit{\textbf{Influence of Length of Short-Term News Sequences on Performance}.} 
To explore the influence of short-term news sequence length on recommendation performance, we conducted experiments to optimize the value of \textit{M}, the number of most recent news items used to model short-term user interests. In our setup, the entire browsing history consists of 50 news items, which we use to model long-term user interests. For short-term interest modeling, we tested a range of sequence lengths from 5 to 30, all of which are naturally shorter than the full history length. Our results, shown in Figure 7, indicate that setting \textit{M}=20 yields the best performance within this range. A sequence length shorter than 20 may not sufficiently capture user preferences, while a longer sequence may weaken the short-term signals by incorporating patterns more representative of long-term user behavior. 

\begin{figure}[htp]
    \centering
    \includegraphics[width=9cm,height=5cm]{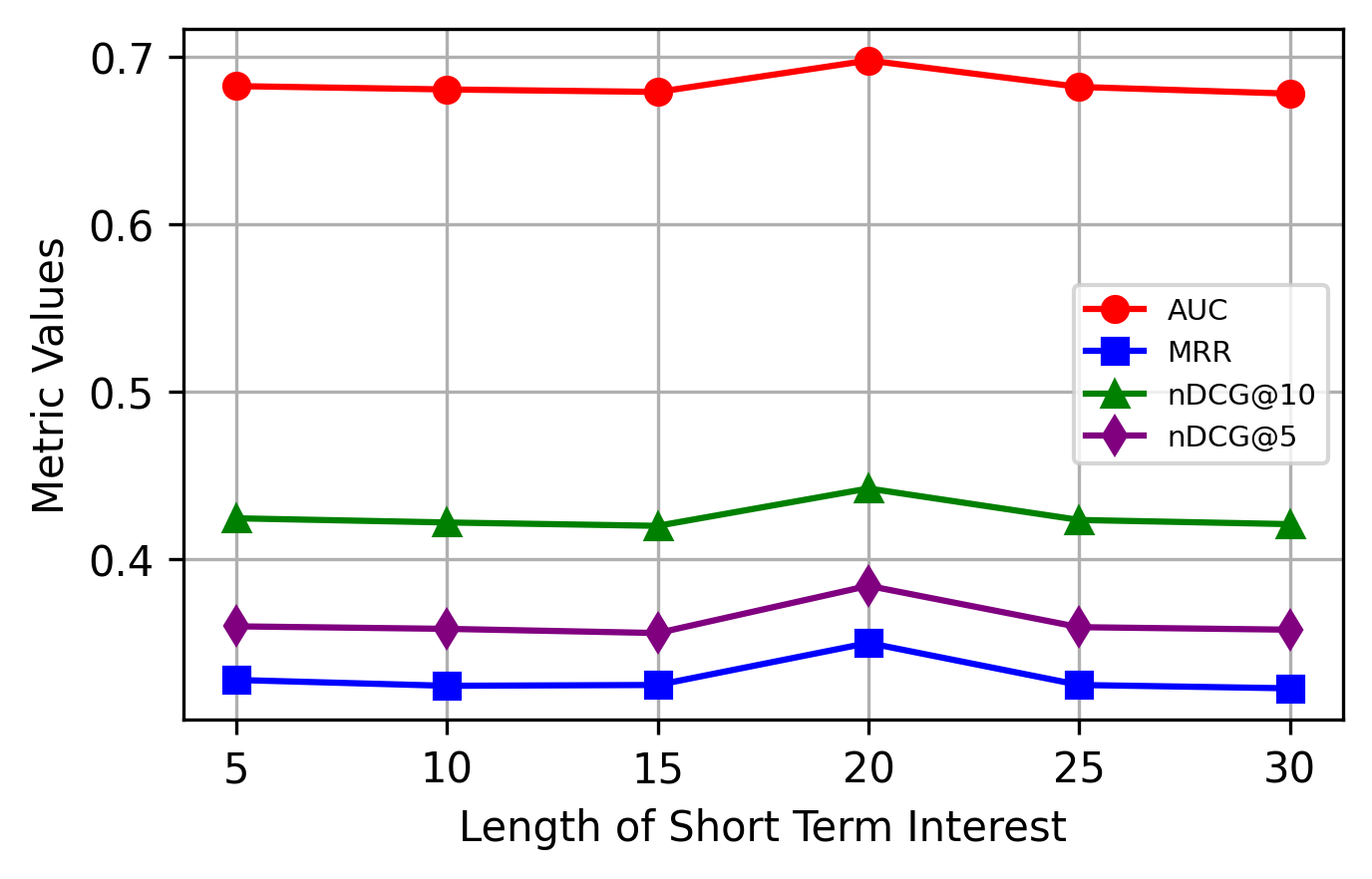}
    \caption{Performance Metrics for Short-Term News Sequence
 Length}
    \label{fig:Figure8}
\end{figure}

\

\noindent \textit{\textbf{Influence of User Group Size on Performance}.} To evaluate the influence of user group size on the performance of the system, we varied the number of users selected per training round and analyzed its impact on the quality of recommendation, as illustrated in Figure 8. The results show that while overall performance remains relatively stable across a range of group sizes, it reaches its peak when the group size is set to 200. Performance tends to improve as the group size increases up to this point, but begins to decline when the group size exceeds 200. This suggests that selecting 200 users per round provides the most effective balance, optimizing the model’s ability to learn from diverse user data.

\begin{figure}[htp]
    \centering
    \includegraphics[width=9cm,height=5cm]{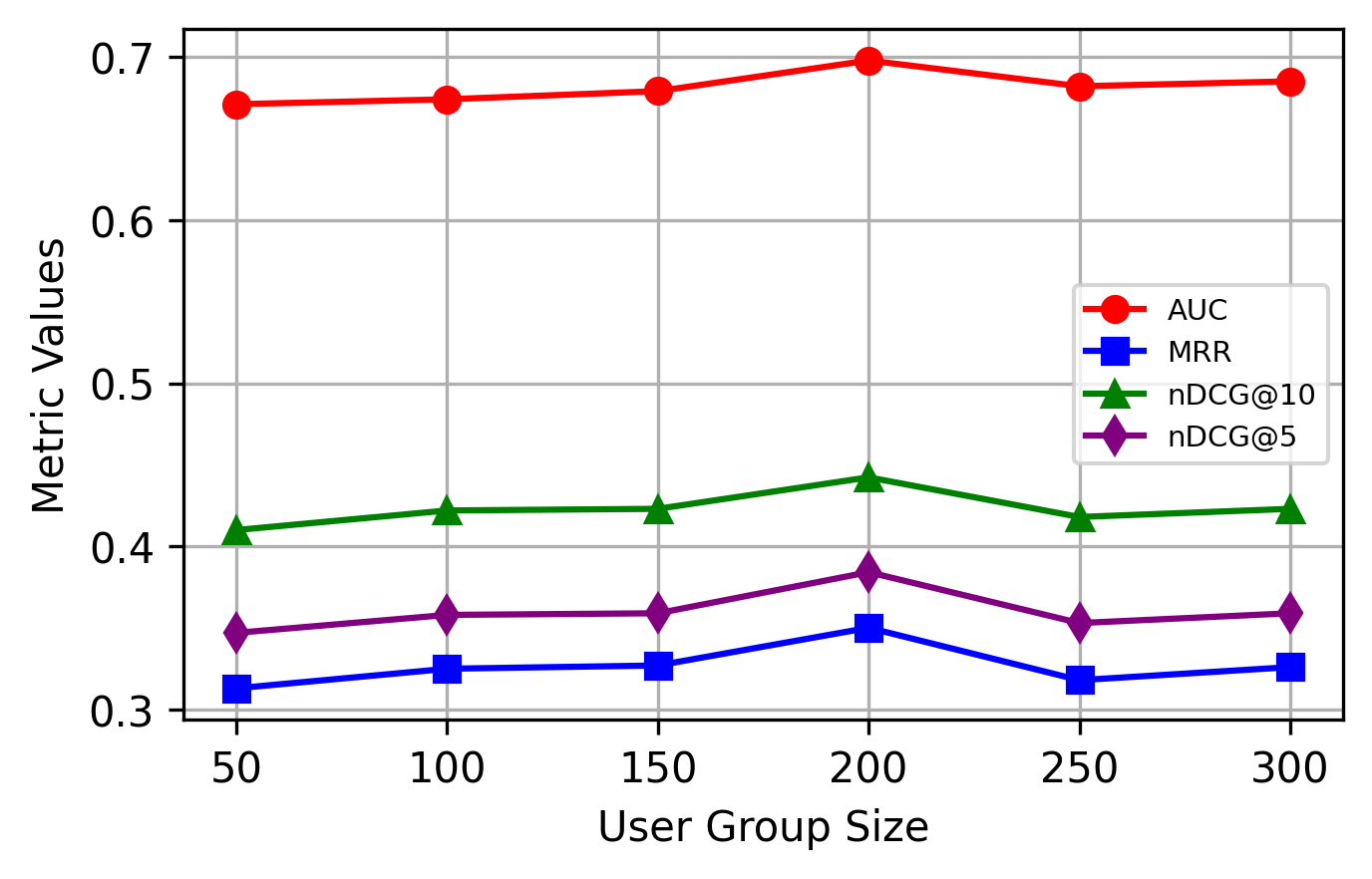}
    \caption{Performance Metrics of User Group Size}
    \label{fig:Figure9}
\end{figure}

\noindent \textit{\textbf{Scalability and Efficiency}.}  To evaluate the scalability of the system, we varied the number of users selected per training round and measured the corresponding training time. As shown in Figure 9, training time generally increases with group size due to the larger aggregated news pool, which elevates both communication and computation overhead. While this increase follows an approximately linear trend up to a group size of 200, beyond this point the slope of the curve rises significantly, indicating a sharp and disproportionate increase in training time. This suggests that larger group sizes (greater than 200) lead to substantially higher resource consumption. Importantly, the group size of 200 not only maintained efficient training time within the linear range but also yielded the highest recommendation accuracy among tested configurations.

\begin{figure}[htp]
    \centering
    \includegraphics[width=9cm,height=5cm]{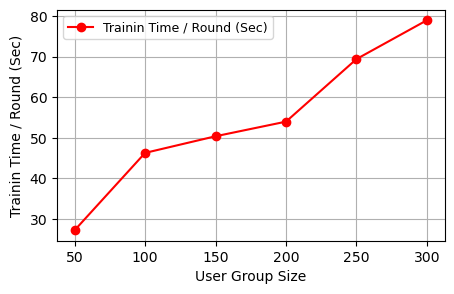}
    \caption{Training Time vs. User Group Size}
    \label{fig:Figure10}
\end{figure}

\section{Conclusion, Limitations, and Future Works}

Our multimodal federated learning framework for news recommendation, Fed-MM-PNR,  integrates both textual and visual features of news while capturing long-term and short-term user interests and preserving user privacy. It demonstrates superior performance over existing systems, with federated learning and secure aggregation algorithms that ensure robust privacy protection without compromising quality. By aligning with existing web infrastructure, this privacy-preserving framework can be adopted by major platforms such as Microsoft and Google News, enabling scalable and decentralized personalized news recommendation systems (PNR).


The primary limitation of this research is the scarcity of standard multimodal datasets containing both textual and visual news information, which constrained the comprehensive evaluation of our model’s effectiveness. We used the MIND dataset and its variants (V-MIND and MM-MIND) that include both the text and cover images of news articles, so we only had one dataset available to train and test our model, limiting our ability to evaluate its generalizability across diverse data sources.

In future work, we plan to enhance privacy by incorporating homomorphic encryption and quantum secure aggregation as additional defense mechanisms. Another promising direction is to extend this approach to other domains beyond news recommendation, such as fashion and e-Commerce recommendation systems, where both multimodal information (e.g., images and text) and federated learning play critical roles in delivering personalized content.

\clearpage

\printbibliography

\end{document}